\newcommand{\CLASSINPUTtoptextmargin}{2cm}
\newcommand{\CLASSINPUTbottomtextmargin}{2cm}
\documentclass[journal]{IEEEtran}
\usepackage{amsmath,amsfonts}
\usepackage{algorithmic}
\usepackage{algorithm}
\usepackage{array}
\usepackage[caption=false,font=normalsize,labelfont=sf,textfont=sf]{subfig}
\usepackage{textcomp}
\usepackage{stfloats}
\usepackage{url}
\usepackage{verbatim}
\usepackage{graphicx}
\usepackage{cite}
\usepackage{balance}

\makeatletter

\providecommand{\tabularnewline}{\\}

\makeatother

\begin{document}
\title{Modelling Scenarios for Carbon-aware Geographic Load Shifting of Compute Workloads}
\author{Wim Vanderbauwhede
        % <-this % stops a space
\thanks{The author is with the School of Computing Science, University of Glasgow, Glasgow, UK}% <-this % stops a space
\thanks{Received 3 September 2025; revised 16 April 2026; accepted 7 June 2026.}
}

\markboth{IEEE Transactions on Sustainable Computing, 2026}%
{Vanderbauwhede: Modelling Scenarios for Carbon-aware Geographic Load Shifting of Compute Workloads}

%\IEEEpubid{0000--0000/00\$00.00~\copyright~2025 IEEE}

\maketitle
\begin{abstract}
We present a linear, statistics based analytical model to evaluate the reductions in $CO_2e$ emissions
resulting from geographic load shifting. The model calculates embodied and operational emissions of data centres participating in geographic load shifting as a function of their load and evaluates the amount of work that can be shifted, and the ensuing change in operational and overall emissions. 
This model is
optimistic as it ignores issues of grid capacity, demand and curtailment.
In other words, real-world reductions will be smaller than the estimates.
However, even with these assumptions, the presented scenarios show
that the realistic reductions from carbon-aware geographic load shifting
are small, of the order of 5\%. This is not enough to compensate the growth in emissions from global data centre expansion.
\end{abstract}


\section{Introduction}

\subsection{Context: Emissions from AI data centres}

\IEEEPARstart{G}reenhouse gas emissions from Information and Communication Technology (ICT) have been estimated at 4\% \cite{knowles2022our,belkhir2018assessing} and are rising steeply. These estimates are from before the rise in popularity of generative AI -- currently the main driver for the growth in emissions from ICT. International consulting firm McKinsey projects a global growth in data centres until 2030 of between 19\% and 27\% annually \cite{mckinsey_data_centre_growth}. Extrapolating their medium scenario of 22\% growth to 2040 corresponds to a 20$\times$ increase in data centre energy consumption, which is far larger than the projected 4$\times$ in \cite{belkhir2018assessing}. 

In 2023, the global electricity demand of data centres was estimated at 55 GW\cite{mckinsey_data_centre_growth,shehabi2024}; that means the projected total power capacity for 2030 will be 1,100 GW and the energy consumption 9,640 TWh/y. For reference, the current world electricity production is about 30,000 TWh/y \cite{owidci2025b}. Furthermore, emissions from power generation are still rising \cite{iea2025}. If these trends persists, AI data centres will become the main contributor to emissions from ICT and to a large fraction of the global carbon budget (see e.g. \cite{vanderbauwhede2025aihype} for more details). It is therefore imperative to reduce emissions from data centre construction and activity. 

\begin{figure}[!t]
        \centering
        \includegraphics[width=8cm]{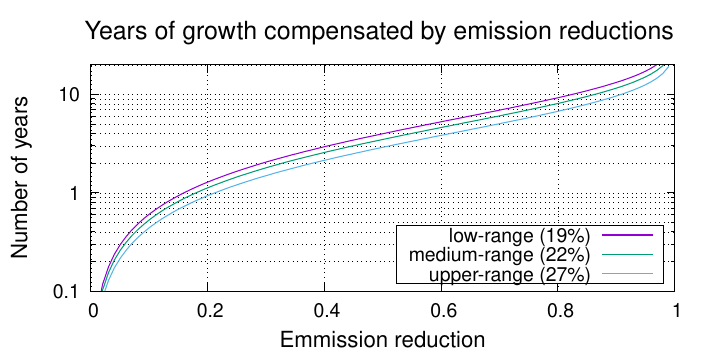}
        
        \caption{Years of growth as per the McKinsey scenarios compensated by emission reductions}
        \label{growth_undoing_reductions}
        \end{figure}

Given exponential growth in data centre power capacity as per the McKinsey scenarios, any approach which reduces yearly data centre emissions by a fixed amount is essentially compensating for a certain time of continued growth. If the approach reduces emissions 10\% per year, this reduction will be undone by the medium-range growth in less than a year. Fig. \ref{growth_undoing_reductions} shows the time it takes for the growth to undo the reduction in emissions (or put positively, the number of years of growth compensated).

This figure shows that for reductions of $<20\%$, the growth will undo the reduction in less that a year. If we can achieve reductions of $>50\%$, we can compensate several years, and that may be enough because economic exponential growth bubbles rarely last more that a decade \cite{morris2008analysis}.

\subsection{Aim and Contributions}

Our aim is to model the reduction in overall data centre emissions resulting from shifting work from a data centre in a high-emission region to one in a low-emission region. 

An important research question largely overlooked in the current research is: even without constraints on the grid, so that the optimal amount of energy can always be delivered anywhere (a "perfect" grid), what is the expected level of emission reductions from deploying these geographic load shifting approaches, assuming realistic parameters for the data centres? %In other words, what is the potential of geographic load shifting?

The contributions of this paper consist of a model to assess the extent of reduction in emissions that can result from geographic load shifting 
and a number of scenarios, both for commercial AI data centres and HPC centres. This model improves on the state of the art in the following ways:

\begin{itemize}
\item Both embodied $CO_2e$ emissions and operational $CO_2e$ emissions are calculated based on up-to-date, detailed and rigorous Life Cycle Analysis models.
\item We combine the calculated emission values into a novel, simple but expressive analytical model that allows to calculate the reduction in emissions arising from geographic load shifting.
\item We consider not only the maximum theoretically possible gains but focus on realistic scenarios.
\end{itemize}

\section{Related work: Carbon-aware computing}

Carbon-aware computing \cite{radovanovic2022carbonaware}
refers to running compute workloads when and where the electricity grid is  powered by renewable energy. This requires shifting work in time and/or space. Time-based shifting is discussed e.g. in \cite{wiesner2021let}; geographic shifting (also known as spatial load shifting or load migration) and spatio-temporal shifting are discussed in detail in this section. The purpose is to reduce emissions from data centre based compute activities in recognition of the steep growth in demand. In this paper we focus on geographic load shifting.

% Lindberg:

Lindberg\emph{ et al.} \cite{LINDBERG2022loadshifting} present a model where data centres shift load
independently of the grid system operators. Relative to previous models for data centre load shifting, their
model improves accuracy and includes more realistic assumptions regarding
the operation of both data centres and the electricity market. Their approach is able to ``reduce
total system carbon emissions by roughly 33\%'' , for
$0.01\leq\epsilon\leq0.2$ where $\epsilon$ is the fraction of the
data centre capacity (load) that can
be shifted per 5-minute time step. This does not mean only that percentage
of load gets effectively shifted: from the graphs in the paper we see that the
average percentage of load shifted is about 17\%.

One major issue with the paper is that it does not list values of the electricity generation carbon intensity (CI) used in the model, nor the locations of the generators
or data centres. It is therefore not possible to directly validate the results.
The actual data are contained in the RTS-GMLC model used in the study,
but this model \cite{8753693} comes with a caveat: ``While projecting
the RTS model onto the southwestern United States enables the use
of geospatially and temporally coincident weather-driven data, it
would be inappropriate to use the projected RTS model to provide insights
into the real-world power system in this location.'' In other words,
the results do not relate directly to real-world scenarios. Furthermore,
the model does not consider data centre embodied carbon nor idle power
consumption or compute load specific shifting constraints. 

% Acun Carbon Explorer acun2023_carbonaware

Acun \emph{et al.} \cite{acun2023_carbonaware} present the Carbon
Explorer framework to analyse the potential for emission reduction
in data centres, including geographic load shifting, and use this
framework to balance trade-offs between operational and embodied based
on geographic location and workload. Their fundamental assumption is that a data centre operator such as Meta can achieve carbon-free operation by buying renewable energy when available. This ignores that renewable generation is overall still a minority fraction
of total generation, and therefore, if the data centre operator buys
all renewable energy it is offloading its emissions onto other users
of the grid.

% Wen-Ting Lin 9712245

%Lin \emph{et al.} \cite{9712245} propose a virtual queue algorithm for job scheduling across data centres in different locations, in other words a spatio-temporal model. The model uses the electricity price and considers both wind and solar. There is no direct information on the CI values used; simulations are based on real-world data from several Australian states, and
%shows traces which indicate the CI lies between 300 and 415 $gCO_{2}e/kWh$.
%The data centre model does take into account idle power but not embodied carbon. It provides no details for active or idle server power consumption or PUE. It shows reductions of between 21\% and 80\% for different scenarios, but these are reductions when compared to a grid without renewables, rather than reductions from spatio-temporal load shifting. The model does not consider compute load specific shifting constraints.

Sukprasert \emph{et al.} \cite{sukprasert2024limitations} conduct a detailed data-driven analysis of the benefits and limitations of carbon-aware spatiotemporal scheduling for cloud workloads. Simulations are based on carbon intensity data from 123 regions. They show that the practical upper bounds of carbon reductions through spatiotemporal workload shifting are currently limited. They find that the benefit of carbon-aware workload scheduling relative to carbon-agnostic scheduling decreases as the grid decarbonises. The paper does not take into account embodied emissions for datacentres nor correlation between sites. It lists a  96\% reduction for 100\% idle capacity at low-CI sites, which means the idle power consumption at high-CI sites is ignored.

% CAFE: Carbon-Aware Federated Learning in Geographically Distributed Data centres 

Bian \emph{et al.} \cite{bian2024carbonawarefl} use carbon-aware geographic
load shifting for federated learning; the paper shows that they can
achieve higher test accuracy for the same carbon budget as existing
approaches. It does not show what reductions in emissions would be
if the test accuracy was kept constant. The model does not consider data centre embodied carbon nor idle power consumption.

% Riepin2025 

Riepin \emph{et al.} \cite{riepin2025spatio} present an approach to spatio-temporal load shifting with the goal of achieving 24/7 Carbon-Free Energy (CFE) matching, meaning that the data centre operator can claim to use 100\% renewable energy. 
%This is similar to \cite{acun2023_carbonaware} and therefore open to the same criticism. 
The paper is thorough in its analysis of the available wind and solar capacity and the correlation between sites. The final result is expressed in terms of cost decrease. It is not possible to tell how much the actual reductions in energy consumption
are as a result of the proposed approach. The model does not consider data centre embodied carbon nor idle power consumption.

% coskun2024 is not 

Coskun \emph{et al.} \cite{coskun2024} propose an
alternative to workload migration: \emph{Conductor} is a framework that coordinates the participation of multiple data centres in demand response, increasing their resilience to operate under power constraints without requiring
any inter-data-centre workload migration. The aim is to provide greater
flexibility in power consumption while improving the ability of 
data centres to meet QoS targets. We can view this as virtual geographic load shifting.

% zheng2020mitigating 

Finally, Zheng \emph{et al.} \cite{zheng2020mitigating} focus on load migration from the Pennsylvania-New Jersey-Maryland interconnection (PMJ) geographical area to California (CAISO), specifically to make use of curtailed renewable energy, and find that geographic load shifting
could absorb up to 62\% of the total curtailment in the CAISO region. If
no additional capacity is installed, shifting work to increase the
load in the California data centres (for a total of 280 MW, 28 data centres of 10 MW) from 50\% to 65\% would result
in a reduction 115 $ktCO_2e/y$. Calculating the baseline from their data gives about 720 $ktCO_2e/y$, assuming that the capacity and load of the data centres in PMJ and CAISO are the same. However, based on their discussion of additional capacity needed, this is not the case: they explain that in the ideal case (all load shifted from PMJ to CAISO), an additional 780 MW at 65\% load would be needed. This means that the amount of load shifted is 550 MW. If we assume
this is the entire load of all PMJ data centres, the baseline is 2.02 $MtCO_2e/y$ (if we assume it is less, then this baseline goes up). In
other words, the best-case reduction without building additional data
centres, 239 $ktCO_e2/y$, amounts to 12\% assuming all work can be shifted. In practice it will be a little less (10\%) because the paper uses an electricity generation carbon intensity of 421 $kgCO_2e/MWh$, whereas the most recent figure is 369 $kgCO_2e/MWh$. 

This is the only work on geographic load shifting that takes into account the embodied carbon of the data centre. Their estimate is based on a figure from \cite{whitehead2015LCA}: ``The study shows that non-operational emissions account for 6.5\% of the total life-cycle climate change impacts of a data centre.'' 
However, that paper expresses all values in Eco-indicator points, and there is not sufficient information in the paper itself to separate the emissions from the other factors contributing to the points. Furthermore, they considerably underestimate the embodied carbon of storage (see e.g. \cite{10.1145/3630614.3630616} for a discussion on SSD and HDD embodied carbon). Therefore, it provides only an approximate estimate for the emissions. Based on this estimate, Zheng \emph{et al.} conclude that ``The embodied GHG emissions of a U.S. data centre therefore amounted to 0.20–0.18 KtCO2e/MW critical power per year during 2016–2019''. This should be contrasted with e.g. the more recent work by \cite{schneider2023} which uses 5 $ktCO_2e/MW$.

\section{Model for emissions reduction of geographic load shifting}

%Our aim is to model the reduction in overall data centre emissions resulting from shifting work from a data centre in a high-emission region to one in a low-emission region. 
The model is not grid-aware: it assumes that there is sufficient power in the target region to power the data centre at full capacity. This is not a fundamental restriction as the available power is only one of the factors that determine the available capacity at the target site. As reduced power availability means less compute capacity, we can model power availability through the load of the data centre; reduced availability of renewables can be modelled through the CI.

The model is conceptually simple: there are only two different types of sites (``high-emission'' and ``low-emission''); we compute the embodied carbon emissions and operational emissions for each site (end-of-life emissions can be ignored, see e.g. \cite{schneider2023,whitehead2015LCA}) assuming that the sites are identical in specification (so we have twin data centres in different locations). The embodied carbon and the operational emissions are calculated using our LCA model \cite{vanderbauwhede2025lifecycleanalysisemissions}. 
Embodied carbon depends on the manufacturing process; operational emissions are based on the power consumption, and the electricity generation carbon intensity of the regions.

We further specify the load of the data centre and the idle power consumption factor, so that we can estimate the power consumption for less than full load. With these parameters, we can calculate the baseline emissions. We also specify which fraction of the work can be moved for which fraction of the time and finally add a factor for the overhead caused by moving the work.

\subsection{Model parameters }

The parameters for the baseline model are:
\begin{description}
\item [{$n_{\textrm{nodes}}$}] number of nodes in data centre. A "node" corresponds to a compute server in a rack, taking into account the overhead of the rack and top-of-rack networking equipment.
\item [{$c_{\textit{emb}}$}] embodied carbon of the data centre ($kgCO_{2}e/y$,
expressed per node). This includes IT equipment, building, cooling and power equipment. It does not include the embodied carbon for the infrastructure outside of the data centre (roads, power lines, communication cables,...).
\item [{$c_{hi}$}] operational carbon emissions of the high-emission site ($kgCO_{2}e/y$,
per node, includes Power Usage Effectiveness (PUE)). This is based on the year-averaged electricity use and electricity generation carbon intensity.
\item [{$c_{lo}$}] operational carbon emissions of the low-emission site ($kgCO_{2}e/y$,
per node, includes PUE)
\item [{$\lambda_{hi}$}] load of high-emission site, $0 \leq \lambda_{hi} \leq 1$. By load we mean the load on the server, i.e. the proportion of the time that it is not idle.
\item [{$\lambda_{lo}$}] load of high-emission site, $0 \leq \lambda_{lo} \leq 1$
\item [{$\gamma$}] idle power consumption as fraction of active power consumption, $0 \leq \gamma \leq 1$ (assumed to be the same for all sites)
\end{description}

The additional parameters for the geographic load shifting model are:
\begin{description}
\item [{$\alpha$}] fraction of workload that can be moved, $0 \leq \alpha \leq 1$.  For example, $\alpha = 0.25$ might represent that only non-latency-sensitive batch jobs can be geo-shifted.
\item [{$\beta$}] fraction of the time that this workload can be moved, $0 \leq \beta \leq 1$. For example, $\beta = 0.5$ could represent that jobs can only be migrated during off-peak hours due to QoS constraints.
\item [{$\eta$}] overhead factor for emissions incurred because of geographic, $0 \leq \eta \ll 1$
load shifting (network emissions, copying of data, ...), on a per-node basis
\end{description}

To simplify the equations, we further define the total emissions from all nodes\begin{equation}
C_{\{emb,hi,lo\}} = c_{\{emb,hi,lo\}}.n_{nodes}~(kgCO_2e/y)
\end{equation}
\subsection{Embodied emissions}

The embodied carbon of the data centre $C_{emb}$ is calculated separately using our model presented in \cite{vanderbauwhede2025lifecycleanalysisemissions}. The model is implemented in the functional programming language Haskell, the source code is available at \cite{hpc_lca_code_wv2025}. We provide here a brief overview of the methodology and sources.

Our model to compute the embodied carbon of server manufacturing is a re-implementation of the model by Boavizta \cite{lorenzini2021digital}. This is a very comprehensive model but as it was published in 2021 we have updated the estimates for various parameters to support more recent technology nodes. %The model includes estimates for the chips and contributions from packaging and assembly, power supplies, motherboard, server enclosure and rack enclosure. 

For the various chips used in the server (CPU, GPU, RAM, SSD) we use the ACT methodology (Architectural Carbon modelling Tool) \cite{10.1145/3470496.3527408}. This methodology uses the electricity consumption of the manufacturing process, the embodied carbon for the materials, and the greenhouse gas potential for the various gases used in production. These parameters are combined with the die size to obtain an estimate for the embodied carbon of the chip. We have updated the parameters that were included without reference in the ACT paper using data from \cite{10.1145/3630614.3630616}, \cite{9372004} and \cite{AnandTech2019}. We also extended the model to include non-integrated GPUs.

In addition, our model also takes into account the embodied emissions of the facility (the data centre building and furnishings, power and cooling infrastructure) and of the networking equipment, based on estimates worked out in \cite{schneider2023}.

\subsection{Emissions from use}

The emissions from use ($c_{hi}$ and $c_{lo}$) are on a per-node bases but do include the overhead of the network infrastructure through $\nu$ and of the cooling etc. through the \emph{PUE} (defined as $\textit{total facility energy} / \textit{IT equipment energy}$).

We define $E_{\textit{node}}$ as the year-average energy consumption of a node, i.e. the average power consumption $P_{\textit{node}}$ multiplied by the time interval. If the node power consumption is expressed in W and the CI is expressed in $kgCO_2e/kWh$, then we have: 
\begin{equation}
E_{\textit{node}} = P_{\textit{node}}\times24\times365/1000~(kWh)
\end{equation}

The emissions from use $C_{\{hi,lo\}}$ are then given by:
\begin{equation}
C_{\{hi,lo\}}(t) = E_{\textit{node}}\cdot n_{\textit{nodes}}\cdot(1+\nu).\textit{PUE}\cdot CI_{\{hi,lo\}}(t)~(kgCO_2e)
\end{equation}

These are the emissions if every node was always working. We combine these with the load $\lambda$ and idle power consumption factor $\gamma$ to get the actual operational emissions.
\begin{equation}\label{Eq:C_op}
C_{op,\{hi,lo\}}(t) = C_{\{hi,lo\}}(t) \cdot (\lambda_{\{hi,lo\}}(t)+\gamma.(1-\lambda_{\{hi,lo\}}(t)))
\end{equation}

The total emissions per site are then
\begin{equation}\label{Eq:C_tot}
C_{tot,\{hi,lo\}}(t) = C_{op,\{hi,lo\}}(t)+C_{\textit{emb}}(t)
\end{equation}

\subsection{Rationale for the model construction}

Our model for geographic load shifting is linear in the sense that the equations do not contain non-linear terms for any of the parameters. The actual model equations are presented and explained in Section \ref{sec:baseline_model} and following. To explain the rationale behind the model construction, it is sufficient to consider the per-site expressions. The embodied emissions are constant; the total emissions are of the form
\begin{equation}
C_{tot}(t)=A\cdot\lambda(t)\cdot C(t)+B\cdot C(t)+D\label{Eq:model_form}
\end{equation}

This is simply a rewrite of Eqs. \ref{Eq:C_op} and \ref{Eq:C_tot} with $A = 1 -\gamma, B =  \gamma, D = C_{\textit{emb}}$, to more clearly show the structure, and with an explicit time parameter.

The load $\lambda(t)$ and the emissions $C(t)$ are statistically independent: the only time-dependent factor in Eq. 3 is the CI. The CI of electricity generation is not influenced by the load and unless the data centre uses temporal shifting, the load is not influenced by the CI either. The load and CI values are time series but the calculation of the total emissions at a given time  does not depend on the values at other times. We can therefore treat $\lambda(t)$ and $C(t)$ as independent random variables and use the mean (expected value) of the distributions, i.e. the time-integrated values of the load and emissions over one year. In general, for statistically independent random variables, the following relations hold between the expected values, regardless of their distributions (see e.g. \cite{kermond2010introduction}):
\begin{equation}
E(XY)=E(X)\cdot E(Y)
\end{equation}

and, with $A$ , $B$ and $D$ constants:
\begin{equation}
E(aX+bY+c)=A\cdot E(X)+B\cdot E(Y)+D
\end{equation}

Therefore, Eq. \ref{Eq:model_form} becomes
\begin{equation}
E(C_{tot}(t))=A\cdot E(\lambda(t))\cdot E(C(t))+B\cdot E(C(t))
\end{equation}

which we write as
\begin{equation}
C_{tot}=A\cdot \lambda\cdot C+B\cdot C+D
\end{equation}

In other words, to obtain the year-averaged emissions for the data centre, there is no need to calculate the emissions at every individual time step, and instead of working with traces we can work with averages. For a more detailed discussion see the Appendix.

\subsection{Baseline emissions model}\label{sec:baseline_model}

We assume that the data centres have excess capacity that could be used for geographic load shifting when there are no peak loads. We take into account the excess capacity using the load parameter $\lambda$ and the idle power consumption parameter $\gamma$. In other words, a load of 50\% means that the data centre is dimensioned
for twice its nominal load. The distinction between high-emission
and low-emission sites is not a strictly geographical one: a given
site could be low-emission or high-emission depending on the energy
mix at a given point in time.
\begin{eqnarray}
C_{b} & = & 2\cdot C_{\textit{emb}}+(\lambda_{hi}+(1-\lambda_{hi})\cdot \gamma)\cdot C_{hi}+\\
 &  & (\lambda_{lo}+(1-\lambda_{lo})\cdot \gamma)\cdot C_{lo}\nonumber
\end{eqnarray}

\subsection{Emissions model including geographic load shifting}\label{sec:gls_model}

Now we assume that we can use some free capacity at the low-emission site by moving load from the high-emission site. We use $\alpha$ and $\beta$ to
express how much of the workload is shifted. The proportion $\alpha$ of the total load that can be shifted expresses constraints on the workload: not all workloads can be shifted, e.g. because they are too large or because of legal, privacy or security considerations. 
The proportion $\beta$ expresses what proportion of the time we can shift load. The rest of the time the emissions are given by the baseline model. For example, it only makes sense to shift load to a region with solar energy generation when that region has  sunshine. As $\alpha$, $\beta$ and $\lambda$ evolve over time, we use yearly averages without loss of generality.
% Need to clarify this because it is the same argument: there is no correlation.
% Use of \beta is explained in 

We have at most $(1-\lambda_{lo})$ free capacity on the low-emission site but the amount of work we can shift is also limited by the freed-up capacity on the high-emission side (because the latter can't exceed the total capacity). When the free load is smaller than the capacity we want to shift, we need to cap $\alpha$ to not exceed free load; when the free load is higher than the load of the high-emission data centre, we need to cap $\alpha$ to not shift more work than is available.
\begin{equation}
\alpha_{\textit{eff}}=\begin{cases}
\alpha & ,\alpha\cdot\lambda_{hi} < (1-\lambda_{lo})\\
\frac{(1-\lambda_{lo})}{\lambda_{hi}} & ,\alpha\cdot\lambda_{hi}\geq (1-\lambda_{lo})
\end{cases}
\end{equation}

By shifting the work, the load on the low site increases from $\lambda_{lo}$
to $\lambda_{lo}+\alpha.\lambda_{hi}$ and reduces on the high
site from $\lambda_{hi}$ to $\lambda_{hi}(1-\alpha)$.

For example, if the load on the high side is $0.9$ and on the low
side $0.7$, and we move \emph{20\%} of the work \emph{50\%} of the
time, then the average load on the high side becomes~$0.9\times(1.0-0.2)=0.72$;
the load on the low side becomes $0.7+0.9\times0.2=0.88$, so that the
total remains 1.6. In other words, 
\begin{equation}
\lambda_{hi}(1-\alpha)+(\lambda_{lo}+\alpha\cdot \lambda_{hi})=\lambda_{hi}+\lambda_{lo}
\end{equation}

The overhead of geographic load shifting is assumed to be proportional
to the amount of work moved and the emissions from use of the nodes:
\begin{equation}
overhead=n_{nodes}\cdot \eta\cdot \alpha\cdot (c_{hi}+c_{lo})
\end{equation}

The intuition for this is that moving work requires additional computations and storage. A detailed analysis for this overhead is presented in \cite{guo2025effect}, which also assumes this proportionality. However, this paper uses Carbon Intensity of Data Transfer (essentially energy/bit transfered) which is a disputed metric, see e.g. \cite{schiencausal} who show that power draw of wired network infrastructure is almost independent of the volume of data traffic, so moving data does not affect the emissions. We will therefore be conservative and use overhead estimates of no more than 1\%, which is on the low side of the interval calculated in \cite{guo2025effect}, $0.5\% < \textrm{migration overhead} < 100\%$. 
%  Consequently, contributing factors to the overhead arising from the geographic load shifting are increase in embodied carbon, which in practice will be mostly additional disk space; and the additional compute required to schedule and manage the load shifting process. There is also a potential overhead in moving a task before it is finished as that requires to interrupt it. If the work is moved on very short timescales (e.g. Lindberg \emph{et al.} assume as short as 5 minutes) then this overhead will not be negligible.

In the above we simplified the discussion by assuming there was a single data centre in the high and low emissions zones. In general, we can have several, and to take this into account we introduce $n_{hi}$ and $n_{lo}$.
With those additional parameters, the constraints become:
\begin{equation}
\alpha_{\textit{eff}}=\begin{cases}
\alpha & ,\alpha\cdot n_{hi}\cdot \lambda_{hi} < n_{lo}\cdot (1-\lambda_{lo})\\
\frac{n_{lo}\cdot (1-\lambda_{lo})}{n_{hi}\cdot \lambda_{hi}} & ,\alpha\cdot n_{hi}\cdot \lambda_{hi}\geq n_{lo}\cdot (1-\lambda_{lo})
\end{cases}
\end{equation}

The final model equation for geographic load shifting becomes
\begin{eqnarray}
C_{\textit{gls}} & = & (n_{hi}+n_{lo})\cdot C_{emb}+\\
 &  & n_{hi}\cdot (\lambda_{hi}(1-\alpha_{\textit{eff}})+\nonumber \\
 &  & (1-\lambda_{hi}(1-\alpha_{\textit{eff}}))\cdot \gamma)\cdot C_{hi}+\nonumber \\
 &  & n_{lo}\cdot (\lambda_{lo}+\alpha_{\textit{eff}}\cdot \lambda_{hi}\cdot n_{hi} /n_{lo}+\nonumber \\
 &  & (1-\lambda_{lo}-\alpha_{\textit{eff}}\cdot \lambda_{hi}\cdot n_{hi}/n_{lo})\cdot \gamma)\cdot C_{lo}+\nonumber \\
 &  & \eta\cdot \alpha_{\textit{eff}}\cdot (n_{hi}C_{hi}+n_{lo}C_{lo})\nonumber
\end{eqnarray}

Taking into account the factor $\beta$, the complete model becomes
\begin{equation}
C = \beta\cdot C_{\textit{gls}}+(1-\beta)\cdot C_b
\end{equation}

If $\beta=0$ we have the baseline model.

\subsection{Emission reductions from geographic load shifting }

The relative reduction in emissions is simply the relative difference
between the baseline and geographic load shifting models:

\begin{equation}
r=\left(C_{b}-C_{ls}\right)/C_{b}
\end{equation}

If the load is the same on both sites, the equation for the ``ideal'' case (no embedded carbon, zero idle, no overhead, move everything all the time) reduces to:

\begin{equation}
r=\begin{cases}
\frac{C_{hi}-C_{lo}}{C_{hi}+C_{lo}} &, \lambda \leq 0.5\\
1-\frac{(2\lambda-1)\cdot C_{hi}+C_{lo}}{\lambda\cdot (C_{hi}+C_{lo})} &, \lambda > 0.5
\end{cases}\label{Eq:reduction_ideal}
\end{equation}

This equation provides the upper limit for achievable reductions in emissions under ideal conditions. Already this provides some insights: if the load is $<0.5$, the achievable reduction does not depend on the load; for the limiting case when the load is one, no reductions are achievable. For example, for a load of 0.5 or less, if we move all work from the US to the UK all the time, (electricity carbon intensity resp. 369 $gCO_2e/kWh$ and 211 $gCO_2e/kWh$ \cite{owidci2025}), the reduction in emissions is $158/580=0.272$, in other words we can never reduce emissions by more than 27\% by doing so.

As the ideal conditions can't exist in reality, the more detailed model, which takes into account embodied carbon, load, idle power consumption, availability and overhead, allows us to make specific predictions of the reductions for concrete scenarios. The model was implemented in Haskell, the source code is available online \cite{gls_code_wv2025}.

\section{Scenarios for geographic load shifting}

\begin{figure*}[!t]
\centering
\includegraphics[width=14cm]{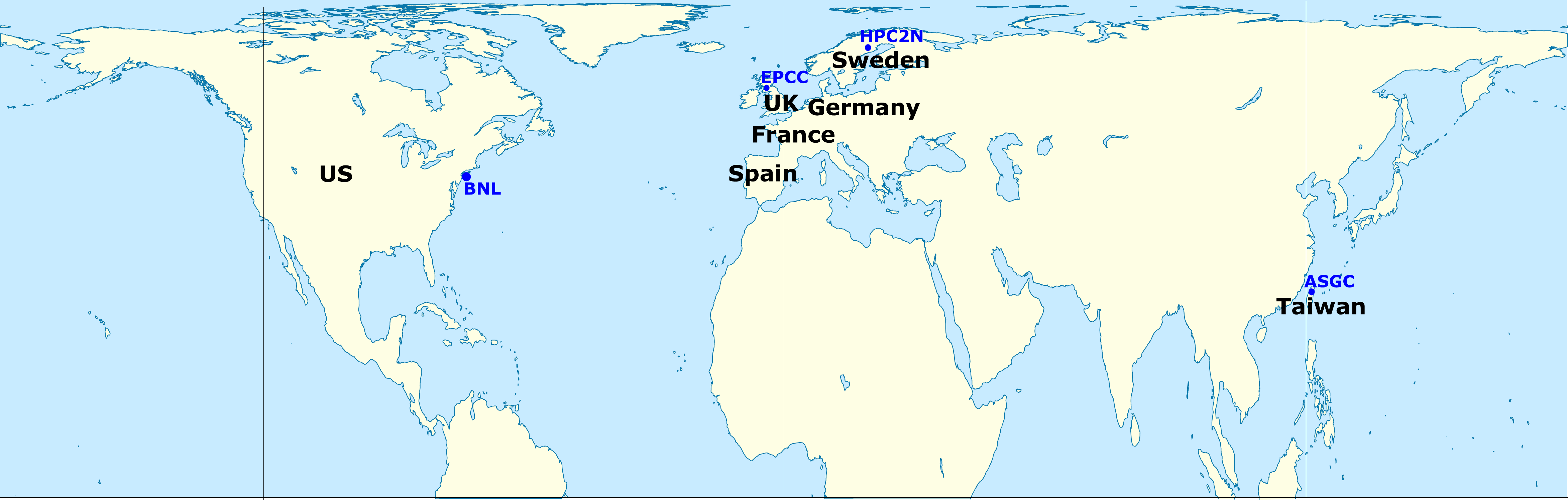}

\caption{Locations used in the scenarios. US, UK, Germany, France and Spain are used in the commercial data centre scenarios, the HPC centres BNL in Brookhaven, US; EPCC in Edinburgh, UK; HPC2N in Umeå, Sweden; ASGC in Taipei, Taiwan are used in the supercomputer scenarios.
}
\label{fig_locations}
\end{figure*}

In this section we apply the above models to different scenarios. We consider scenarios for a commercial data centre attempting to optimise emissions through use of solar energy and wind energy and for HPC centres offloading work to centres in low-emission regions. The locations used are shown in Fig. \ref{fig_locations}. The countries are used in the commercial data centre scenarios, the HPC centres in the HPC scenarios.

\subsection{Scenarios for commercial data centres}

In these scenarios, summarised in Table \ref{tab:scenarios-AI},
we want to assess the relative reduction in emissions resulting
from geographic load shifting for AI cloud data centres. We base our scenarios on the same assumptions
as in the work by Lindberg \emph{et al.} \cite{LINDBERG2022loadshifting}, which is one of the papers proposing geographic load shifting to reduce carbon emissions, but which approaches the problem from a grid cost optimisation perspective, with the data centre grid load as an opaque parameter.
Lindberg \emph{et al.} assume four identical data centres in different locations,
with different but unspecified electricity carbon intensity. The maximum
load is 300 MW and the nominal load 250 MW. The absolute figure is
not important as we want to investigate the relative reduction in emissions,
but 250 MW is a realistic capacity for a very large hyperscale data
centre (hyperscale data centres of more than twice this capacity are currently being proposed \cite{ravenscraig}).

There is no detail on the internals of the data centre in the work by Lindberg \emph{et al.}.
We assume an AI data centre where the nodes are GPU servers similar
to the Nvidia DGX-A100. Each node consumers 4,550 W when active \cite{nvidia2024},
30\% of that when idle \cite{amdepyc2019,speccpu2020,Yang_2024,zhang2024improving}; the data centre PUE is taken to be 1.16, representative for a hyperscale data centre (but probably on the low side: most data centres have a higher PUE). We set the number of server nodes so that in total the data centre consumes at most 300 MW. In other words, the load for 250 MW is $\lambda=0.83$ (250/300) and we can move a fraction $\alpha=0.2$ (50 MW out of 250 MW).

Our embodied carbon model yields an estimate for this configuration of 5,730 $kgCO_2e/y$ per node.  This assumes a 4-year useful life and takes into account the data centre infrastructure.  This corresponds to 0.45 $ktCO_2e/MW/y$, which is between the estimates of \cite{zheng2020mitigating} and \cite{schneider2023}.

Table \ref{tab:scenarios-AI} shows the parameters used and
the results.

\begin{table*}
\caption{\label{tab:scenarios-AI}
Parameters for geographic load shifting to optimise use of solar energy (Scenario 1)\protect \\
and wind energy (Scenario 2)}
\centering
\begin{tabular}{|c|c|c|}
\hline 
%%%%%%%%%
\textbf{AI Data Centre Parameters} & \textbf{Scenario 1 (Solar)} & \textbf{Scenario 2 (Wind)}\tabularnewline
\hline
\hline
$n_n$ & 56,840 & 56,840\tabularnewline
\hline
$n_{hi}$ & 2 & 2\tabularnewline
\hline
$n_{lo}$ & 2 & 2\tabularnewline
\hline
Embodied carbon $c_{em}\, (kgCO_2e/y)$ & 2,066 & 2,066\tabularnewline
\hline
Operational emissions, high-CI $c_{hi}\, (kgCO_2e/y)$ & 18,978 & 17,451\tabularnewline
\hline
Operational emissions, low-CI $c_{lo}\, (kgCO_2e/y)$ & 1,896 & 509\tabularnewline
\hline
$\lambda_{hi}$ & 0.83 & 0.83\tabularnewline
\hline
$\lambda_{lo}$ & 0.83 & 0.83\tabularnewline
\hline
$\gamma$ & 0.30 & 0.30\tabularnewline
\hline
$\alpha$ & 0.20 & 0.20\tabularnewline
\hline
$\beta$ & 0.52 & 0.63\tabularnewline
\hline
$\eta$ & 0.00 & 0.00\tabularnewline
\hline
\emph{overhead} ($tCO_2e/y$) & 0 & 0\tabularnewline
\hline
Embodied ($tCO_2e/y$) & 469,692 & 469,692\tabularnewline
\hline
Baseline ($tCO_2e/y$) & 2,565,724 & 2,273,088\tabularnewline
\hline
Geographic load shifting ($tCO_2e/y$) & 2,447,922 & 2,151,753\tabularnewline
\hline
Emission reduction (\%) & 4.6\% & 5.3\%\tabularnewline
\hline

%%%%%%%%%

\end{tabular}
\end{table*}

\subsubsection{Scenario 1: Periodic load shifting to optimise use of solar energy}

In this scenario, we start by assuming that for eight hours a day, three of the four
data centres use predominantly solar power and for the rest of the
day predominantly fossil fuels. To make the example
concrete, we assume that the other locations are the UK, the US and Germany, with average carbon intensities of resp. 211, 369 and 344 $gCO_2e/kWh$\cite{owidci2025}. We assume the fourth data centre is located in France and uses nuclear when there is no sun. We assume for simplicity that the carbon intensity during solar powered operation is the same as the average CI. This is acceptable as according to the
IPCC \cite{ipccAR5annexIII}, solar power has an average CI of 41 and France has an average CI of 44 $gCO_2e/kWh$. According to \cite{en14113143solarPV}, the CI varies between 6.5 and 108 $gCO_2e/kWh$ depending on location and technology
used, with most sites between 20 and 50 $gCO_2e/kWh$.

We further assume that the CI per country is the time-weighted average of the CI of predominantly solar and  predominantly fossil fuel generation (the energy mix typically includes non-fossil fuel sources such as nuclear and wind energy):
\begin{equation}
\mathit{CI}_{\textit{avg}}=(8\times\mathit{CI}_{\textit{sun}}+16\times\mathit{CI}_{\textit{fossil}})/24
\end{equation}
 
We calculate the CI for the period with predominantly fossil fuel generation from the average CI for the country and the CI for solar energy generation. For example for the UK: 
\begin{eqnarray}
\mathit{CI}{}_{\mathit{fossil,UK}} & = & (24\times\mathit{CI}{}_{\mathit{avg,UK}}-8\times\mathit{CI}_{\textit{sun}})/16 \\
 & = & 295~(gCO_2e/kWh)\nonumber
\end{eqnarray}

The average of the thus calculated CI from predominantly fossil fuel generation over the
three countries is 448 $gCO_2e/kWh$; the predominantly solar generation is assumed to be 41 $gCO_2e/kWh$ as per the IPCC \cite{ipccAR5annexIII}. 

\begin{figure*}[!t]
\centering
(a)~\includegraphics[width=9cm]{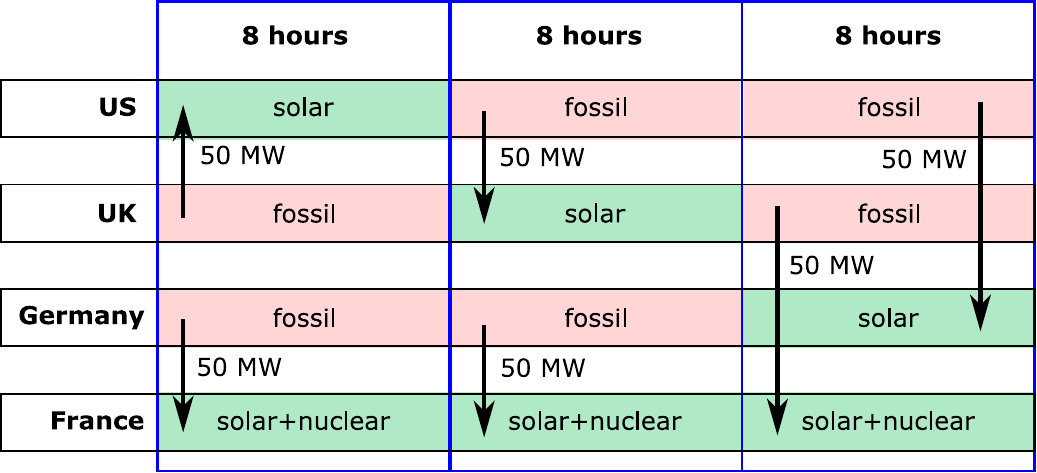}\\~\\
(b)~\includegraphics[width=5cm]{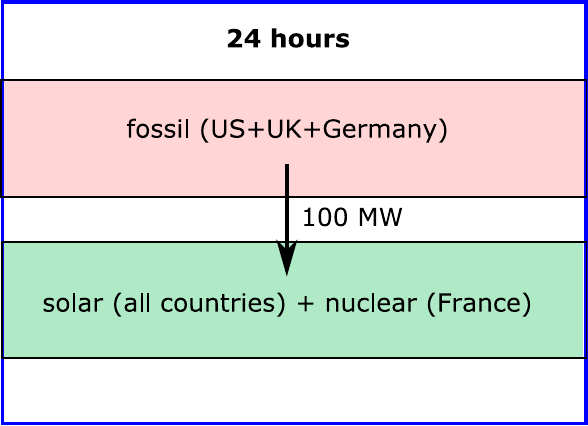}
\caption{Solar scenario, ideal assumptions (two sites are low-CI in any 8-hour period). 100 MW can be shifted.}
\label{fig_solar_1}
\end{figure*}

As illustrated in Fig. \ref{fig_solar_1}, at any point in time there will always be two data centres with low CI and two with high CI. Fig. \ref{fig_solar_1}(a) shows the generation for every data centre in every 8-hour timespan; Fig. \ref{fig_solar_1}(b) shows that we can reorder this so that it corresponds to an equivalent load shifting model where 50 MW is moved from a high CI location to low CI one (so 100 W in total) all of the time, even though in reality we will move the load 2/3 of the time and to different sites. In other words, we can model this 4-site scenario using an equivalent 2-site scenario as used in our model. We use the average CI for the three high-emission regions for $CI_{hi}$. This is an optimal scenario as we can always move the maximum load. 

As formulated, the scenario is overly optimistic as it assumes that
in all three countries the sun will shine eight hours a day year round or 2,920 hours of sunshine per year. For this scenario, in the UK, the average number of hours of sunshine per year is 1,524 hours \cite{metofficeSunshine}; in the US it is 2,627 hours (average over all states) \cite{noaaSunshine} and in Germany it is 1,665 hours \cite{DeWetterSunshine}. For France, there is of course no such correction, so the average number of hours is $(h_{US}+h_{UK}+h_D+3.h_F)/6=0.83$. We correct for this by reducing the fraction of the time that the workload can be moved, $\beta$, by this factor.
Furthermore, the scenario assumes that there is at all times a data centre in a region where the sun shines. If we have two data centres in Europe and one in the US, this is not the case. (We exclude the one in France as it is always low-CI because of the nuclear power.) 
Excluding Hawaii, US time ranges from UTC-8 to UTC-4. So we will assume the US data centre to be on UTC-6; Europe is on UTC+1. 

\begin{figure*}[!t]
\centering
\includegraphics[width=9cm]{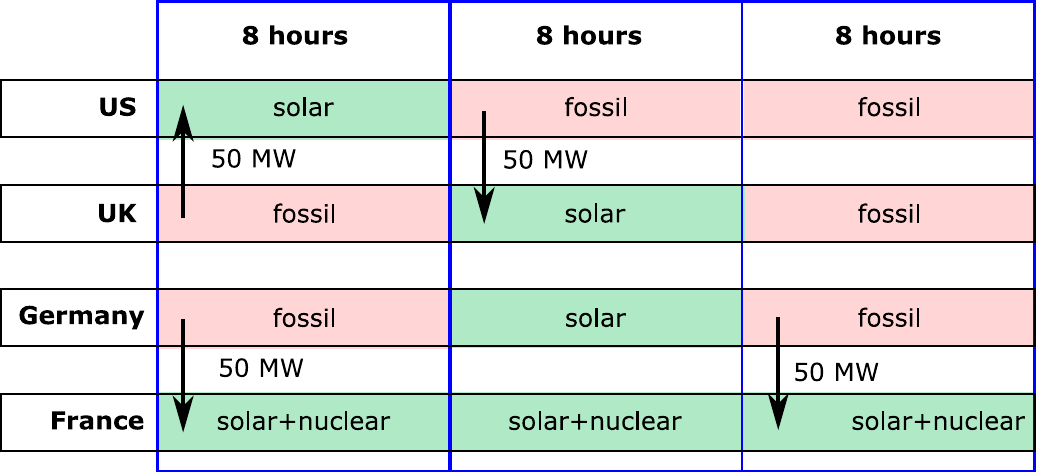}
\caption{Solar scenario, more realistic assumptions (two, three or one low-CI sites in an 8-hour period). Only (100+50+50)/3 MW can be shifted.}
\label{fig_solar_2}
\end{figure*}

As illustration, in Fig. \ref{fig_solar_2} we simplify to no overlap between the US, the UK and Europe (i.e. we assume the entire US is on UTC-8 and Europe is on UTC). Then there is one 8-hour period where one out of three data centres is in a low emissions region, one with two and another with none, as illustrated in Fig. \ref{fig_solar_2}. In the first period, we can move 100 MW; in the second and third period, we can only move 50 MW, so in total we can only move 200 MW rather than 300 MW. In practice, there will be 2 hours overlap between the US and the UK and 1 hour non-overlap between Germany and the UK, so that the final factor is not 2/3 but 5/8. We account for this further correcting $beta$ to  $0.52$. 
% A more realistic assumption is that the entire US is on UTC-5 and Europe on UTC+1
%XXXX XX X X ______ _ ___ ____ __
%____ __ X X XXXXXX _ ___ ____ __
%____ __ _ X XXXXXX X ___ ____ __
%XXXX XX X X XXXXXX X XXX XXXX XX
%6*100
%1*50
%1*0
%6*50
%1*100
%9*50
%100*6+1*50+1*0+6*50+1*100+9*50
%100*8+50*16
%5/8

\subsubsection{Scenario 2: Periodic load shifting to optimise use of wind energy }

\begin{figure*}[!t]
\centering
\includegraphics[width=11cm]{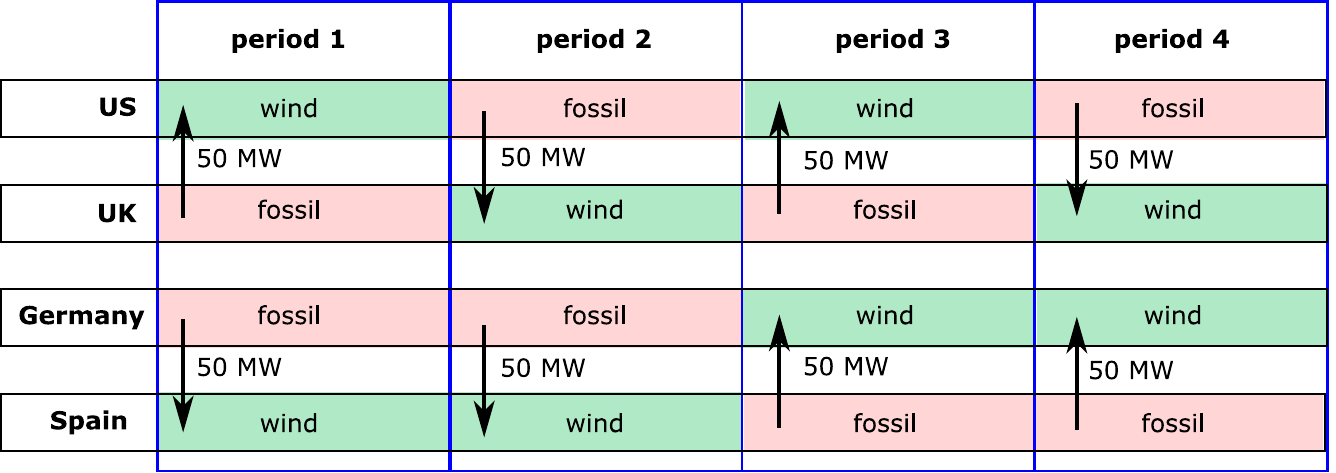}\\~\\
\caption{Wind scenario, ideal assumptions (wind load factor 50\% for all sites, no correlation). 100 MW can be shifted.}
\label{fig_wind_1}
\end{figure*}

In this scenario we assume that the four data centres are in the US, the UK, Germany and Spain (the choice of Spain is because after Germany, it is the EU country with the largest wind power capacity installed). We start by assuming that on average there is always enough wind in the regions of two out of four data centres (Fig. \ref{fig_wind_1}). This means that we can always move the load from the other two, so this amounts to the same situation as for Scenario 1, and we can move 100 MW between sites. For the wind generation CI we use 11 $gCO_2e/kWh$ as per the IPCC \cite{ipccAR5annexIII}. According to \cite{alsaleh2019wind,GARCIATERUEL2022wind,LI2021wind,ozoemena2018wind}, reported values vary between 10.3 and 45.2 $gCO_2e/kWh$. The average CI from predominantly fossil fuel generation over the four countries, calculated as above, is 531 $gCO_2e/kWh$.

\begin{figure*}[!t]
\centering
\includegraphics[width=10cm]{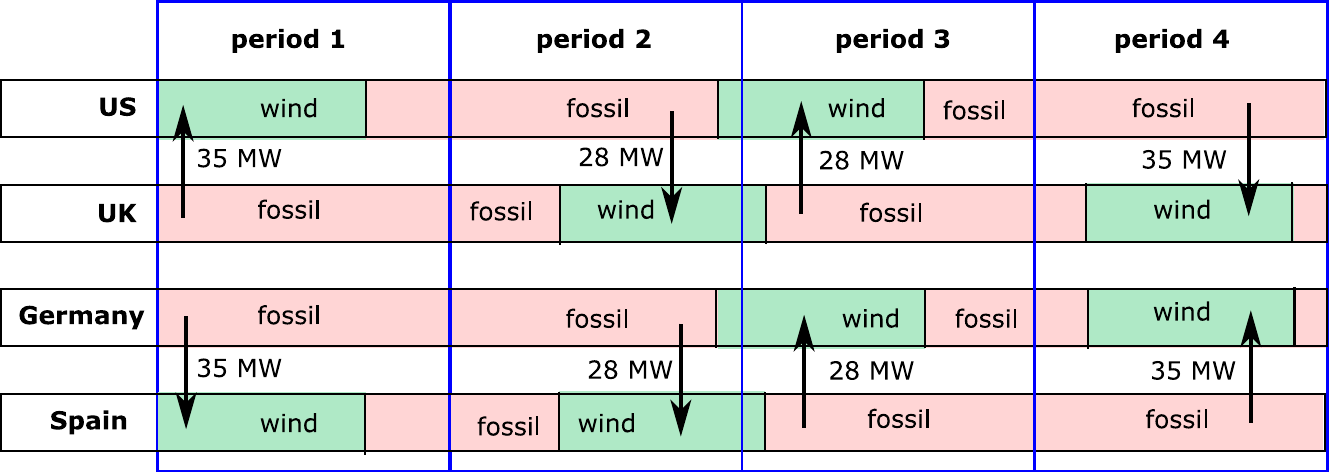}
\caption{Wind scenario, more realistic assumptions (load factor 35\% and correlation 10\% ). Only 63 MW can be shifted.}
\label{fig_wind_2}
\end{figure*}

As formulated, the scenario is again overly optimistic as it assumes there will always be enough wind to move 100 MW. In practice, load factors are not 50\% but between 30\% and 40\% or lower \cite{ALBATAYNEH2025windload,STAFFELL2014windload}; we will therefore reduce $\beta$ to $0.7$. Furthermore, there tends to be a correlation in the weather patterns across Europe, so that it will not always be possible to move all the load. Maps for the Pearson correlation for hourly wind capacity within Europe are presented in \cite{riepin2025spatio}. Based on those values and assuming no correlation with the US, using the average correlation between the UK, Germany and Spain reduces the effectiveness by 10\%, so we finally set $\beta=0.63$. This is illustrated in Fig. \ref{fig_wind_2}. In this example, in periods 1 and 4 there is no correlation, so load shifting is reduced from 50 MW to 35 MW by the wind load factor; the windy episodes in periods 2 and 3 are overlapping by 20\% (full correlation would mean complete overlap), leading to a further reduction to 28 MW. So in total, only 63 MW can be shifted instead of 100 MW.

\subsubsection{Discussion of Commercial Data Centre Scenarios}

As we can see from Table \ref{tab:scenarios-AI}, for both scenarios, which are still quite optimistic, the reduction in emissions is around 5\% (4.6\% and 5.3\% respectively for the solar and wind scenarios). Several factors conspire to limit the gains:

\begin{itemize}
\item the carbon intensity of any current renewable technology is not zero. 
\item the CI of predominantly fossil fuel generation is lower than the worst
case (100\% coal generation is 820 $gCO_2e/kWh$; 100\% gas is 490 $gCO_2e/kWh$,
\cite{ipccAR5annexIII}; most data centres are in the EU and the US which both have CI lower than this (207 $gCO_2e/kWh$ resp. 369 $gCO_2e/kWh$) 
\item the important contribution of embodied carbon in the low-emission
case,
\item the idle power consumption is not zero and 
\item in realistic scenarios we can't move sufficient load
to use up all excess capacity in the low-emission region all the time. 
\end{itemize}

\begin{figure*}[!t]
\centering
\includegraphics[width=15cm]{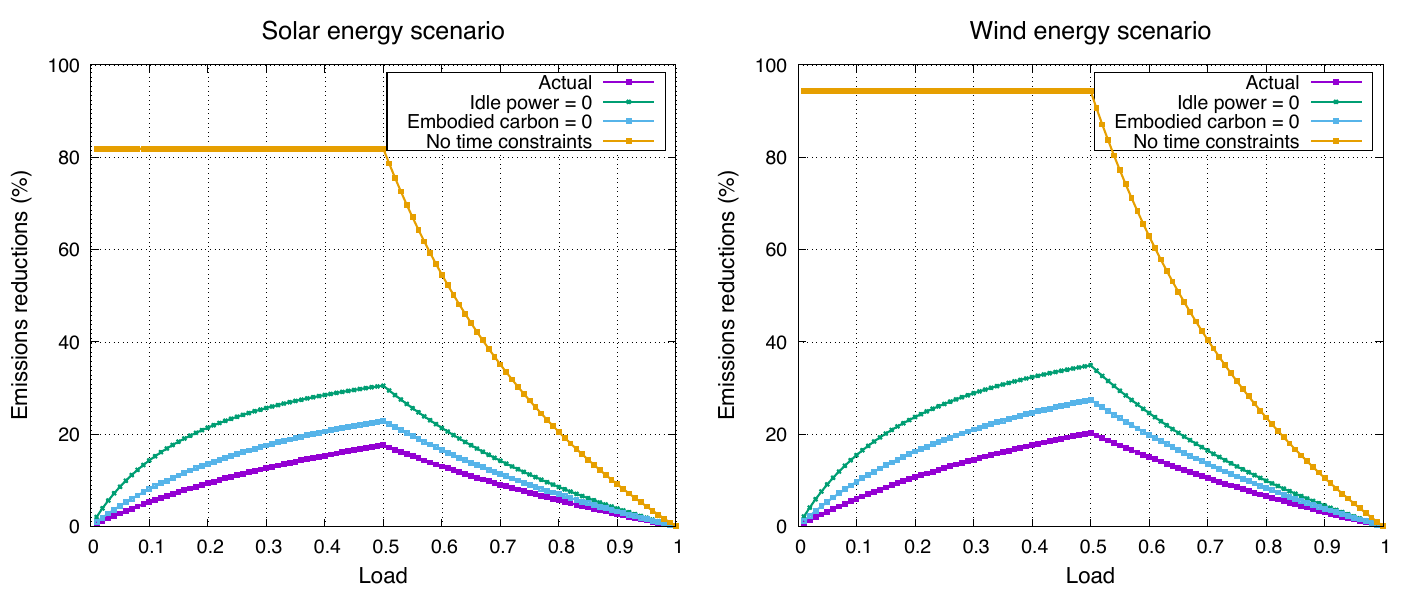}
\caption{Emissions reduction as a function of load for the AI data centre scenarios. The figure illustrates the effect of load $\lambda$, idle power $\gamma$, embodied emissions $c_{em}$ and time constraints $\beta$. }
\label{fig_effect_load}
\end{figure*}

To explore these factors, in Fig. \ref{fig_effect_load} we plot the emission reductions as a function of the load. We move as much work as possible. The set of curves shows the combined effect of each factor: "Idle power = 0" removes the effect of the idle power consumption; "Embodied carbon = 0" removes the embodied carbon contribution; "No time constraints" does both and ignores the time constraints on the renewables generation. The kink in the figure is the result of the constraint on $\alpha$, which in these scenarios is triggered when the load exceeds 0.5. The intuition is that for loads $< 0.5$, the amount of work that can be shifted is limited by the available work in the high-emission zone, whereas for loads $> 0.5$, the limit is the free capacity in the low-emission zone. The reason why the reduction is independent of the load is that all work is moved to the low-emission zone, so that effectively the reduction is given by the second case in Eq. \ref{Eq:reduction_ideal}, $(1-CI_{lo})/(CI_{hi}+CI_{lo})$

Our model ignores the actual grid capacity. The low-emission location is only low emission if the additional energy required by the shifting of the workloads does not exceed the curtailed renewables generation. If data centre operators deploying carbon-aware computing move more work than the curtailed capacity then non-renewable generators will switch on and the result will be an increase in electricity generation carbon intensity. Using grid-aware computing, less work would be shifted to avoid exceeding the curtailed capacity. In practice, curtailment is quite low. According to the IEA \cite{iea2023}, it is between 1.5\% and 4\% in most large renewable energy markets, so there is not all that much scope for moving data centre workloads at scale. 

\subsection{Supercomputer centre scenarios}

In this section we consider a different type of scenario, that of moving workloads between supercomputers. HPC centres are built for a notional capacity and their workloads do not depend strongly on market factors. In a network such as CERN's Worldwide LHC Computing Grid (WLCG), load could be moved between locations with high and low emissions if spare capacity is available.

We model this using actual values for the ASGC HPC centre in Taiwan and the HPC2N centre in Sweden, CI 642 $gCO_2e/kWh$ resp. 36 $gCO_2e/kWh$, as well as BNL in the US and EPCC in the UK, CI resp. 369 $gCO_2e/kWh$ and 211 $gCO_2e/kWh$ (CI values from \cite{owidci2025}).

%Embodied carbon and emissions from use for the servers are calculated using our own state-of-the-art models for HPC centres \footnote{https://codeberg.org/wimvanderbauwhede/low-carbon-computing/src/branch/master/LCA-model-equations}. 

We assume a node consisting of two AMD EPYC 9754 HT with 1 TB RAM and 0.5 TB NVME SSD, with a node power consumption of 1.2 kW. Our embodied carbon model yields an estimate for this configuration of 444 $kgCO_2e/y$. This assumes a 4-year useful life and takes into account the data centre infrastructure. This corresponds to 0.37 $ktCO_2e/MW/y$ and is again between the existing estimates. 

Table \ref{tab:scenarios-HPC} shows the parameters used and
the results.

\begin{table*}
\caption{\label{tab:scenarios-HPC}
Scenario 1: excess capacity is 50\%, maximum possible amount of work is moved, no overhead;\\
Scenario 2: as Scenario 1 but excess capacity is 20\%, overhead 1\%;\\
Scenario 3: as Scenario 2 but only move 12.5\% of the work;\\
Scenario 4: as Scenario 2 but move between BNL (US) and EPCC (UK);\\
Scenario 5: Scenario 3 with CI of Scenario 4;\\
Values for emissions are in $kgCO_2e/y$}

\centering
\begin{tabular}{|c|c|c|c|c|c|}
\hline 
%%%%%%%%
\textbf{HPC Centre Parameters} & \textbf{Scenario 1} & \textbf{Scenario 2} & \textbf{Scenario 3} & \textbf{Scenario 4} & \textbf{Scenario 5}\tabularnewline
\hline
\hline
$n_n$ & 100 & 100 & 100 & 100 & 100\tabularnewline
\hline
$n_{hi}$ & 1 & 1 & 1 & 1 & 1\tabularnewline
\hline
$n_{lo}$ & 1 & 1 & 1 & 1 & 1\tabularnewline
\hline
Embodied carbon $c_{em}\, (kgCO_2e/y)$ & 444 & 444 & 444 & 444 & 444\tabularnewline
\hline
Operational emissions, high-CI $c_{hi}\, (kgCO_2e/y)$ & 10,831 & 10,831 & 10,831 & 3,879 & 3,879\tabularnewline
\hline
Operational emissions, low-CI $c_{lo}\, (kgCO_2e/y)$ & 390 & 390 & 390 & 1,304 & 1,304\tabularnewline
\hline
$\lambda_{hi}$ & 1.00 & 0.80 & 0.80 & 0.80 & 0.80\tabularnewline
\hline
$\lambda_{lo}$ & 0.50 & 0.80 & 0.80 & 0.80 & 0.80\tabularnewline
\hline
$\gamma$ & 0.30 & 0.30 & 0.30 & 0.30 & 0.30\tabularnewline
\hline
$\alpha$ & 1.00 & 1.00 & 0.25 & 1.00 & 0.25\tabularnewline
\hline
$\beta$ & 1.00 & 1.00 & 0.50 & 1.00 & 0.50\tabularnewline
\hline
$\eta$ & 0.00 & 0.01 & 0.01 & 0.01 & 0.01\tabularnewline
\hline
\emph{overhead} ($tCO_2e/y$) & 0 & 3 & 1 & 1 & 1\tabularnewline
\hline
Embodied ($tCO_2e/y$) & 89 & 89 & 89 & 89 & 89\tabularnewline
\hline
Baseline ($tCO_2e/y$) & 1,197 & 1,054 & 1,054 & 534 & 534\tabularnewline
\hline
Geographic load shifting ($tCO_2e/y$) & 832 & 910 & 982 & 500 & 517\tabularnewline
\hline
Emission reduction (\%) & 30.5\% & 13.6\% & 6.8\% & 6.5\% & 3.3\%\tabularnewline
\hline

%%%%%%%%

\end{tabular}
\end{table*}

\subsubsection{Scenario 1}

In this scenario we model moving workloads from the ASGC HPC centre in Taiwan, which has a very high electricity carbon intensity and a PUE of 1.62, to the HPC2N centre in Sweden, which has a very low electricity carbon intensity and a PUE of 1.03. This is the most optimistic case. From discussion with HPC facility operators, we assume an average load of 80\%.  However, for the first scenario we assume HPC2N has 50\% free capacity for the entire year and ASGC has a load of 100\%. We further assume that we would like to move as much of the work for as much of the time as possible from the ASGC cluster to HPC2N. These are very optimistic assumptions. We assume no overhead in moving the work. In this case, the reduction in emissions is 30.5\% or about 365 $tCO_2e/y$. We start with this scenario to demonstrate that only highly unrealistic scenarios result in high reductions.

\subsubsection{Scenario 2}

In this scenario we assume that both ASGC and HPC2N have an average load of 80\%, which is a realistic load for HPC centres, and that we can move as much work as possible (i.e. so that HPC2N is 100\% loaded). We also assume an overhead of 1\% for moving the work. There are no data for this overhead, but even if it was zero, this would not change the figures much. This results in a reduction in emissions of 13.6\%. 

\subsubsection{Scenario 3}

In practice, it will not be possible to move all of the work all of the time. Many HPC workloads are limited in particular by large data sets and large volumes of data produced, which would make the workload movement impractical, or by software or data set licensing or data privacy issues which prohibit work movement. In this scenario, we assume that we can move a quarter of the workload for half of the time. This amounts to moving 12.5\% instead of 20\%. The resulting reduction in emissions is 6.8\%. 

\subsubsection{Scenario 4}

Moving workloads from sites with very high CI to ones with very low CI is of course optimal, but most HPC centres are neither in very high nor very low emission regions. We therefore look at moving workloads between a relatively high- and low-emission site, BNL in the US (PUE 1.35) and EPCC in the UK (PUE 1.1). Using the other parameters from Scenario 2, this results in a reduction in emissions of 6.5\%.

\subsubsection{Scenario 5}

Finally, in what is probably the most realistic scenario, we use Scenario 3 with the CI of Scenario 4. In other words, we move a realistic amount of work between sites with a realistic difference in CI. This results in a reduction in emissions of 3.3\%. 
\begin{figure}[!h]
        \centering
        \includegraphics[width=7.5cm]{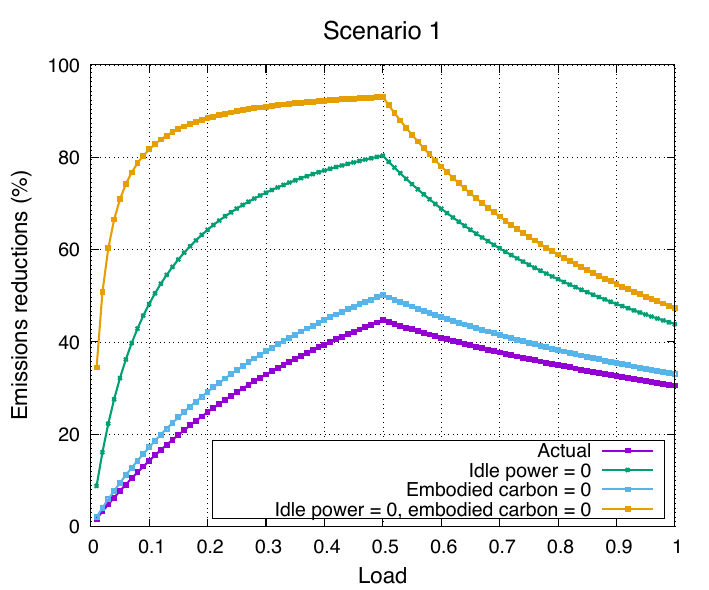}
        \caption{Emissions reduction as a function of load for HPC centre Scenario 1, illustrating the effect of load $\lambda$, idle power $\gamma$ and embodied emissions $c_{em}$. }
        \label{fig_effect_load_HPC_opt}
        \end{figure}
\subsubsection{Discussion of HPC Scenarios}

\begin{figure*}[!t]
        \centering
        \includegraphics[width=13.5cm]{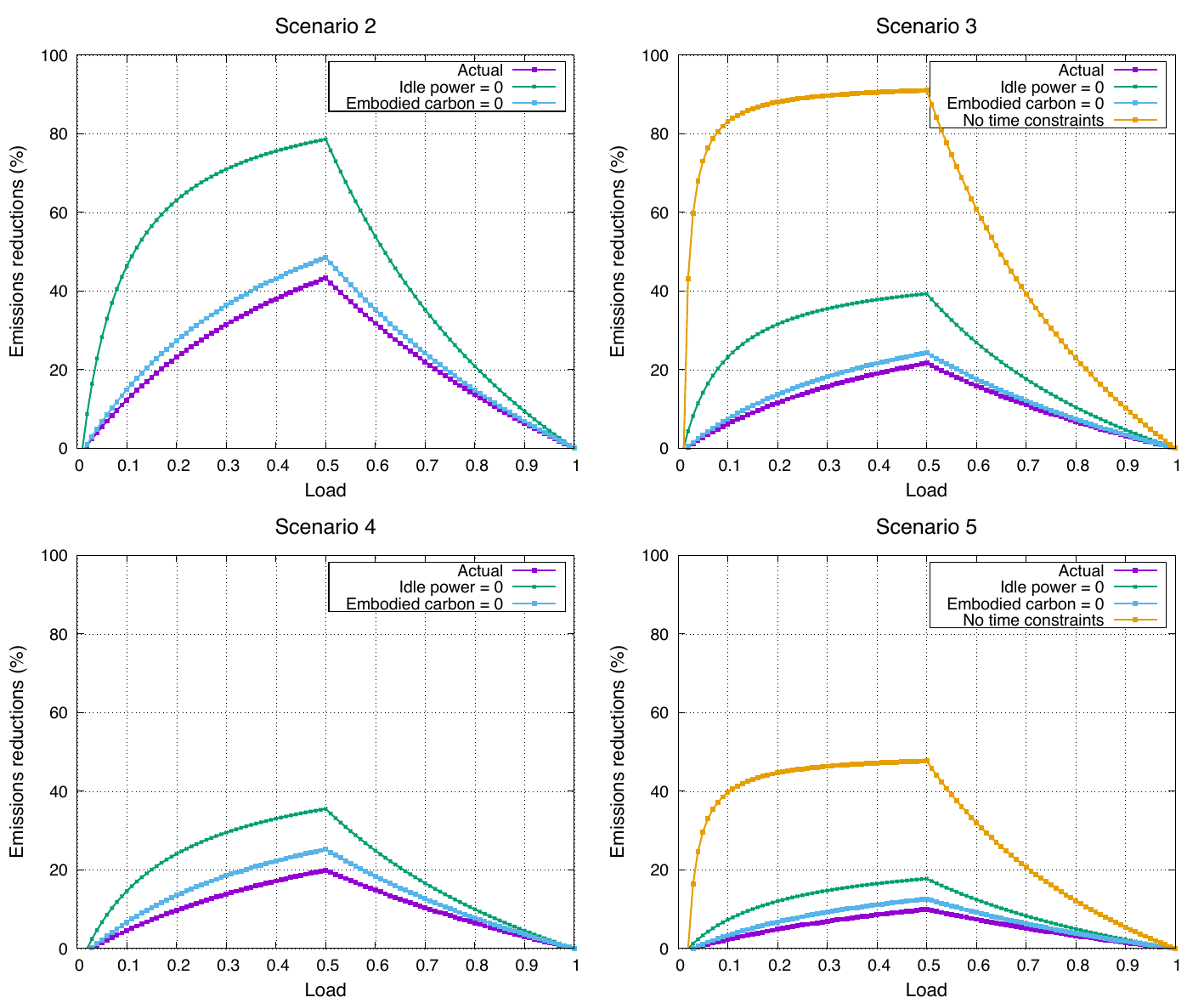}
        \caption{Emissions reduction as a function of load for the HPC centre scenarios. The figure illustrates the effect of load $\lambda$, idle power $\gamma$, embodied emissions $c_{em}$ and time constraints $\beta$. }
        \label{fig_effect_load_HPC}
        \end{figure*}

For Scenario 1, Fig. \ref{fig_effect_load_HPC_opt} shows the effect of load $\lambda$, idle power $\gamma$, embodied emissions $c_{em}$. In this scenario, $\lambda_{lo}=0.5$ and we sweep $\lambda_{hi}$. Fig. \ref{fig_effect_load_HPC} explores the effect of load $\lambda$, idle power $\gamma$, embodied emissions $c_{em}$ and load flexibility $\beta$. For Scenarios 2 and 4, $\beta=1$ and therefore the "No time constraints" case is redundant. Comparing Scenario 2/4 to Scenario 3/5 shows the impact of the flexibility in moving work; comparing Scenario 2/3 to Scenario 4/5 shows the impact of the CI differential.

These supercomputer scenarios show that, even if the data centres don't provision excess capacity and therefore don't incur excess embodied carbon, geographic load shifting only results in large reductions if the free capacity at the low-emission HPC centre is very high, the high-emission site has very high emissions and the low-emission site very low ones, and the overhead for moving the workloads is very small (Scenario 1). Unless there is complete flexibility in moving workloads, i.e. we can move the work all the time, the most likely obtainable emission reductions are around five percent. But many scientific computing workloads are limited in particular by large data sets and large volumes of data produced, which would make the workload movement impractical. Other techniques such as frequency downscaling for I/O-limited workloads \cite{jackson2023archer} or heat reuse \cite{LJUNGDAHL2022117671} can yield much higher reductions, and should therefore be prioritised. Reducing embodied carbon through server life extension will become increasingly important as the grid decarbonises. 
        
\section{Conclusion}

Based on realistic scenarios for commercial AI data centres and HPC centres, and a linear analytical model that takes into account key server and data centre parameters, we have simulated the reduction in emissions from geographic load shifting. 

We conclude that achievable reductions in emissions are small (typically less than 5\%) for realistic scenarios for both commercial data centres and HPC centres, even though we optimistically assume a perfect grid and ignore any performance penalties. Much larger reductions in emissions are needed to counter the current growth in global data centre capacity. 

A possible strategy to achieve such reductions is demand reduction, which could be achieved 
%through pricing mechanisms but requires incentives to change business models. An obvious way would be to include 
by including the ICT sector in emissions cap-and-trade schemes such as the EU Emissions Trading System.

\section*{Acknowledgement}

We thank Dr. Lauritz Thamsen for his thorough review and helpful suggestions which led to a much improved manuscript.

\bibliographystyle{IEEEtran}
\bibliography{TSUSC-2025-09-0250-final}

@misc{vanderbauwhede2025aihype,
    Author = {Vanderbauwhede, Wim},
    Month = {01},
    Title = {{The real problem with the AI hype}},
    year = {2025},
    url={https://limited.systems/articles/the-real-problem-with-AI/},
    urldate={2025-05-08}
}

@article{vanderbauwhede2025lifecycleanalysisemissions,
  title = {Life Cycle Analysis for Emissions of Scientific Computing Centres},
  author = {Wadenstein, Mattias and Vanderbauwhede, Wim},
  year = {2025},
  month = aug,
  journal = {The European Physical Journal C},
  volume = {85},
  number = {8},
  pages = {913},
  issn = {1434-6052},
  doi = {10.1140/epjc/s10052-025-14650-8}
}

@misc{owidci2025,
    author={Ember},
    year= 2025, 
    note={Energy Institute - Statistical Review of World Energy (2024) - with major processing by Our World in Data. },

    title = {{Carbon intensity of electricity generation - Ember and Energy Institute}},
    url         = {https://ourworldindata.org/grapher/carbon-intensity-electricity},
    urldate     = {2025-05-30}
}

@misc{owidci2025b,
    author={Ember},
    year= 2025, 
    note={Energy Institute - Statistical Review of World Energy (2024) - with major processing by Our World in Data. },
        title = {{Electricity generation from other renewables, excluding bioenergy – Ember}},
    url         = {https://ourworldindata.org/grapher/electricity-prod-source-stacked},
    urldate     = {2025-05-30}
}

@article{LINDBERG2022loadshifting,
    title = {Using geographic load shifting to reduce carbon emissions},
    journal = {Electric Power Systems Research},
    volume = {212},
    pages = {108586},
    year = {2022},
    issn = {0378-7796},
    doi = {https://doi.org/10.1016/j.epsr.2022.108586},
    url = {https://www.sciencedirect.com/science/article/pii/S0378779622006757},
    author = {Julia Lindberg and Bernard C. Lesieutre and Line A. Roald},
}

@misc{nvidia2024,
    title={{Energy and Power Efficiency for Applications on the Latest NVIDIA Technology}},
    booktitle={{GTC 2024}},
    author = { Gray, A.},
    month={Mar},    
    year={2024},
    url={https://developer-blogs.nvidia.com/wp-content/uploads/2024/10/Energy-Efficiency-GTC.pdf},
    urldate={2025-01-20}
}

@misc{amdepyc2019,
    title={{AMD EPYC 7742 Benchmarks and Review Simply Peerless}},
    author={Kennedy, P.},
    year={2019},
    month={dec},
    booktitle={{ServeTheHome}},
    url={https://www.servethehome.com/amd-epyc-7742-benchmarks-and-review-simply-peerless/},
    urldate={2025-01-20}
}

@misc{speccpu2020,
    title={{SPEC CPU 2017 Integer Speed Result}},
    author={{Standard Performance Evaluation Corporation}},
    url={http://spec.org/cpu2017/results/res2020q2/cpu2017-20200413-21925.pdf},
    year={2019},
    urldate={2025-01-20}
    
}

@inproceedings{Yang_2024,
   title={Accurate and Convenient Energy Measurements for GPUs: A Detailed Study of NVIDIA GPU's Built-In Power Sensor},
   url={http://dx.doi.org/10.1109/SC41406.2024.00028},
   DOI={10.1109/sc41406.2024.00028},
   booktitle={Proc. SC24},
   author={Yang, Zeyu and Adamek, Karel and Armour, Wesley},
   year={2024},
   month=nov, pages={1-17} 
   }

@inproceedings{zhang2024improving,
  title={Improving {GPU} energy efficiency through an application-transparent frequency scaling policy with performance assurance},
  author={Zhang, Yijia and Wang, Qiang and Lin, Zhe and Xu, Pengxiang and Wang, Bingqiang},
  booktitle={Proc. EuroSys2024},
  pages={769--785},
  year={2024}
}

@Article{en14113143solarPV,
AUTHOR = {Ziemi\'{n}ska-Stolarska, Aleksandra and Pietrzak, Monika and Zbici\'{n}ski, Ireneusz},
TITLE = {Application of LCA to Determine Environmental Impact of Concentrated Photovoltaic Solar Panels - State-of-the-Art},
JOURNAL = {Energies},
VOLUME = {14},
year = {2021},
NUMBER = {11},
ARTICLE-NUMBER = {3143},
URL = {https://www.mdpi.com/1996-1073/14/11/3143},
ISSN = {1996-1073},
}

@article{GARCIATERUEL2022wind,
title = {Life cycle assessment of floating offshore wind farms: An evaluation of operation and maintenance},
journal = {Applied Energy},
volume = {307},
pages = {118067},
year = {2022},
issn = {0306-2619},
doi = {https://doi.org/10.1016/j.apenergy.2021.118067},
url = {https://www.sciencedirect.com/science/article/pii/S0306261921013520},
author = {Anna Garcia-Teruel and Giovanni Rinaldi and Philipp R. Thies and Lars Johanning and Henry Jeffrey},
}

@article{LI2021wind,
title = {Life cycle assessment and life cycle cost analysis of a 40 MW wind farm with consideration of the infrastructure},
journal = {Renewable and Sustainable Energy Reviews},
volume = {138},
pages = {110499},
year = {2021},
issn = {1364-0321},
doi = {https://doi.org/10.1016/j.rser.2020.110499},
url = {https://www.sciencedirect.com/science/article/pii/S1364032120307851},
author = {Qiangfeng Li and Huabo Duan and Minghui Xie and Peng Kang and Yi Ma and Ruoyu Zhong and Tianming Gao and Weiqiong Zhong and Bojie Wen and Feng Bai and Arun K. Vuppaladadiyam},
}

@article{alsaleh2019wind,
  title={Comprehensive life cycle assessment of large wind turbines in the US},
  author={Alsaleh, Ali and Sattler, Melanie},
  journal={Clean Technologies and Environmental Policy},
  volume={21},
  pages={887--903},
  year={2019},
  publisher={Springer}
}

@article{ozoemena2018wind,
  title={Comparative LCA of technology improvement opportunities for a 1.5-MW wind turbine in the context of an onshore wind farm},
  author={Ozoemena, Matthew and Cheung, Wai M and Hasan, Reaz},
  journal={Clean Technologies and Environmental Policy},
  volume={20},
  number={1},
  pages={173--190},
  year={2018},
  publisher={Springer}
}

@article{STAFFELL2014windload,
title = {How does wind farm performance decline with age?},
journal = {Renewable Energy},
volume = {66},
pages = {775-786},
year = {2014},
issn = {0960-1481},
doi = {https://doi.org/10.1016/j.renene.2013.10.041},
url = {https://www.sciencedirect.com/science/article/pii/S0960148113005727},
author = {Iain Staffell and Richard Green},
}

@article{ALBATAYNEH2025windload,
    title = {Wind farm capacity factor forecasting: An Australian case study},
    journal = {Energy Nexus},
    volume = {18},
    pages = {100422},
    year = {2025},
    issn = {2772-4271},
    doi = {https://doi.org/10.1016/j.nexus.2025.100422},
    url = {https://www.sciencedirect.com/science/article/pii/S2772427125000634},
    author = {Aiman Albatayneh and Ragheb AbuAlRous and Merlinde Kay and Ramez Abdallah and Adel Juaidi and Amos García-Cruz and Francisco Manzano-Agugliaro},
}

@article{radovanovic2022carbonaware,
  title={Carbon-aware computing for datacenters},
  author={Radovanovi{\'c}, Ana and Koningstein, Ross and Schneider, Ian and Chen, Bokan and Duarte, Alexandre and Roy, Binz and Xiao, Diyue and Haridasan, Maya and Hung, Patrick and Care, Nick and others},
  journal={IEEE Transactions on Power Systems},
  volume={38},
  number={2},
  pages={1270--1280},
  year={2022},
  publisher={IEEE}
}

@inproceedings{bian2024carbonawarefl,
  title={{CAFE: Carbon-Aware Federated Learning in Geographically Distributed Data Centers}},
  author={Bian, Jieming and Wang, Lei and Ren, Shaolei and Xu, Jie},
  booktitle={Proceedings of the 15th ACM International Conference on Future and Sustainable Energy Systems},
  pages={347--360},
  year={2024}
}

@inproceedings{acun2023_carbonaware,
    author = {Acun, Bilge and Lee, Benjamin and Kazhamiaka, Fiodar and Maeng, Kiwan and Gupta, Udit and Chakkaravarthy, Manoj and Brooks, David and Wu, Carole-Jean},
    title = {Carbon Explorer: A Holistic Framework for Designing Carbon Aware Datacenters},
    year = {2023},
    isbn = {9781450399166},
    publisher = {Association for Computing Machinery},
    address = {New York, NY, USA},
    url = {https://doi.org/10.1145/3575693.3575754},
    doi = {10.1145/3575693.3575754},
    location = {Vancouver, BC, Canada},
    booktitle = {Proc. ASPLOS 2023},
    series = {ASPLOS 2023}
}

@inproceedings{ipccAR5annexIII,
    author={S. Schlömer  and  T. Bruckner and L. Fulton and E. Hertwich and A. McKinnon and D. Perczyk and J. Roy and R. Schaeffer and R. Sims and P. Smith and R. Wiser}, 
    year={2014},
    title={{Annex III: Technology-specific cost and performance parameters.}},
    booktitle={{Climate Change 2014: Mitigation of Climate Change. Contribution of WG III to the 5th Assessment Report of the IPCC}},
    publisher={Cambridge University Press}, 
    location={Cambridge, United Kingdom},
}

@misc{metofficeSunshine,
    title={{UK temperature, rainfall and sunshine time series}},
    author={{UK Met Office}},
    year={2025},
url=  {https://www.metoffice.gov.uk/research/climate/maps-and-data/uk-temperature-rainfall-and-sunshine-time-series},
    urldate={2025-05-08}

}

@misc{DeWetterSunshine,
title={{Deutschlandwetter im Jahr 2022}},
author={{Deutsche Wetterdienst}},
    year={2022},
    url={https://www.dwd.de/DE/presse/pressemitteilungen/DE/2022/20221230_deutschlandwetter_jahr2022_news.html},
 urldate={2025-05-08}
}

@misc {noaaSunshine,
title={{WMO Climate Normals}},
author={{NOAA National Centers for Environmental Information}},
year={2025},
url={https://www.ncei.noaa.gov/products/wmo-climate-normals},
urldate={2025-06-03}
}

@inproceedings{jackson2023archer,
author = {Jackson, Adrian and Simpson, Alan and Turner, Andrew},
title = {Emissions and energy efficiency on large-scale high performance computing facilities: ARCHER2 UK national supercomputing service case study},
year = {2023},
url = {https://doi.org/10.1145/3624062.3624269},
doi = {10.1145/3624062.3624269},
booktitle = {Proc. SC23},
pages = {1866-1870},
location = {Denver, CO, USA},
}

@article{LJUNGDAHL2022117671,
title = {A decision support model for waste heat recovery systems design in Data Center and High-Performance Computing clusters utilizing liquid cooling and Phase Change Materials},
journal = {Applied Thermal Engineering},
volume = {201},
pages = {117671},
year = {2022},
issn = {1359-4311},
doi = {https://doi.org/10.1016/j.applthermaleng.2021.117671},
url = {https://www.sciencedirect.com/science/article/pii/S1359431121010966},
author = {V. Ljungdahl and M. Jradi and C. Veje}
}

@misc{schneider2023,
    title={{Quantifying Data Center Scope 3 GHG Emissions to Prioritize Reduction Efforts}},
    author={ Lin,P. and Bunger , R. and Avelar, V.},
    year={2023},
    month={may},
    url={https://www.se.com/ww/en/download/document/SPD_WP99_EN/},
    urldate={2025-06-04}
}

@misc{gls_code_wv2025,
    title={{
A model for the achievable reduction in emissions from carbon-aware geographic load shifting
}},
    author={Vanderbauwhede, W.},
    year={2025},
    url={https://codeberg.org/wimvanderbauwhede/carbon-aware-geographic-load-shifting-model},
    urldate={2025-08-18}
}

@misc{hpc_lca_code_wv2025,
  author       = {Wadenstein, Mattias and
                  Vanderbauwhede, Wim},
  title        = {Model for Life Cycle Analysis for Emissions of
                   Scientific Computing Centres
                  },
  month        = aug,
  year         = 2025,
  publisher    = {Zenodo},
  version      = {v1.1},
  doi          = {10.5281/zenodo.16894210},
  url          = {https://doi.org/10.5281/zenodo.16894210},
}

@misc{lorenzini2021digital,
  title={Digital \& environment: How to evaluate server manufacturing footprint, beyond greenhouse gas emissions?},
  author={Lorenzini, R.},
  booktitle={Boavizata Blog},
  year={2021},
  month={Nov},
  publisher={Boavizta},
  url={https://boavizta.org/en/blog/empreinte-de-la-fabrication-d-un-serveur},
  urldate={2025-01-20}
}

@inproceedings{10.1145/3470496.3527408,
author = {Gupta, Udit and Elgamal, Mariam and Hills, Gage and Wei, Gu-Yeon and Lee, Hsien-Hsin S. and Brooks, David and Wu, Carole-Jean},
title = {ACT: designing sustainable computer systems with an architectural carbon modeling tool},
year = {2022},
isbn = {9781450386104},
publisher = {Association for Computing Machinery},
address = {New York, NY, USA},
doi = {10.1145/3470496.3527408},
booktitle = {Proceedings of the 49th Annual International Symposium on Computer Architecture},
pages = {784–799},
numpages = {16},
location = {New York, New York},
series = {ISCA '22}
}

@INPROCEEDINGS{9372004,
  author={Garcia Bardon, M. and Wuytens, P. and Ragnarsson, L.-A. and Mirabelli, G. and Jang, D. and Willems, G. and Mallik, A. and Spessot, A. and Ryckaert, J. and Parvais, B.},
  booktitle={2020 IEEE International Electron Devices Meeting}, 
  title={{DTCO including Sustainability: Power-Performance-Area-Cost-Environmental score (PPACE) Analysis for Logic Technologies}}, 
  year={2020},
  volume={},
  number={},
  pages={41.4.1-41.4.4},
  doi={10.1109/IEDM13553.2020.9372004}}

@article{10.1145/3630614.3630616,
author = {Tannu, Swamit and Nair, Prashant J.},
title = {The Dirty Secret of SSDs: Embodied Carbon},
year = {2023},
issue_date = {October 2023},
publisher = {Association for Computing Machinery},
address = {New York, NY, USA},
volume = {3},
number = {3},
url = {https://doi.org/10.1145/3630614.3630616},
doi = {10.1145/3630614.3630616},
journal = {SIGENERGY Energy Inform. Rev.},
month = oct,
pages = {4–9},
numpages = {6}
}

@misc{ravenscraig,
    title={{Scotland's Ravenscraig proposed for £4bn hyperscale data centre}},
    booktitle={{The Engineer}},
    author = {{Mark Allen Group}},
    month={Jun},    
    year={2025},    
    url={https://www.theengineer.co.uk/content/news/ravenscraig-site-proposed-for-4bn-hyperscale-data-centre},
    urldate={2025-06-27}
}

@misc{AnandTech2019,
    title={{SK Hynix Details DDR5-6400}},
    booktitle={{AnandTech}},
    author = { Shilov, A.},
    month={Feb},    
    year={2019},
    url={https://www.anandtech.com/show/13999/sk-hynix-details-its-ddr56400-dram-chip},
    urldate={2025-01-20}
}

@inproceedings{wiesner2021let,
  title={Let's wait awhile: How temporal workload shifting can reduce carbon emissions in the cloud},
  author={Wiesner, Philipp and Behnke, Ilja and Scheinert, Dominik and Gontarska, Kordian and Thamsen, Lauritz},
  booktitle={Proceedings of the 22nd International Middleware Conference},
  pages={260--272},
  year={2021}
}

@article{riepin2025spatio,
  title={Spatio-temporal load shifting for truly clean computing},
  author={Riepin, Iegor and Brown, Tom and Zavala, Victor M},
  journal={Advances in Applied Energy},
  volume={17},
  pages={100202},
  year={2025},
  publisher={Elsevier}
}

@misc{iea2023,
    title={{Renewable Energy  Market Update -- Outlook for 2023 and 2024}},
    author = {{International Energy Agency}},
    month={Jun},    
    year={2023},
    urldate={2025-01-20},
	url={https://www.iea.org/reports/renewable-energy-market-update-june-2023/will-more-wind-and-solar-pv-capacity-lead-to-more-generation-curtailment}
}

@misc{iea2025,
    title={{Electricity 2025 -- Analysis and forecast to 2027}},
    author = {{International Energy Agency}},
    month={Oct},    
    year={2025},
    urldate={2026-01-07},
	url={https://iea.blob.core.windows.net/assets/7c671ef6-2947-4e87-beea-af0e1288e1d7/Electricity2025.pdf}
}

@INPROCEEDINGS{coskun2024,
  author={Acun, Fatih and Paschalidis, Ioannis Ch. and Coskun, Ayse K.},
  booktitle={2024 IEEE 15th International Green and Sustainable Computing Conference (IGSC)}, 
  title={Conductor: A Collaboration Framework for Multi-Data-Center Demand Response}, 
  year={2024},
  volume={},
  number={},
  pages={22-28},
  doi={10.1109/IGSC64514.2024.00013}}

@article{zheng2020mitigating,
  title={Mitigating curtailment and carbon emissions through load migration between data centers},
  author={Zheng, Jiajia and Chien, Andrew A and Suh, Sangwon},
  journal={Joule},
  volume={4},
  number={10},
  pages={2208--2222},
  year={2020},
  publisher={Elsevier}
}

@article{schiencausal,
author = {Schien, Daniel and Shabajee, Paul and Krug, Louise and McSorley, Greg and Preist, Chris},
title = {Causal allocation of fixed impacts in product systems: Assessing the effect of data demand on network energy consumption},
journal = {Journal of Industrial Ecology},
volume = {n/a},
number = {n/a},
pages = {},
year={2025},
month={Jul},
doi = {https://doi.org/10.1111/jiec.70057},
url = {https://onlinelibrary.wiley.com/doi/abs/10.1111/jiec.70057},
eprint = {https://onlinelibrary.wiley.com/doi/pdf/10.1111/jiec.70057}
}

@ARTICLE{8753693,
  author={Barrows, Clayton and Bloom, Aaron and Ehlen, Ali and Ikäheimo, Jussi and Jorgenson, Jennie and Krishnamurthy, Dheepak and Lau, Jessica and McBennett, Brendan and O’Connell, Matthew and Preston, Eugene and Staid, Andrea and Stephen, Gord and Watson, Jean-Paul},
  journal={IEEE Transactions on Power Systems}, 
  title={The IEEE Reliability Test System: A Proposed 2019 Update}, 
  year={2020},
  volume={35},
  number={1},
  pages={119-127},
  doi={10.1109/TPWRS.2019.2925557}}

@misc{mckinsey_data_centre_growth,

    title={{AI power: Expanding data center capacity to meet growing demand}},
    author = {{McKinsey \& Company}},
    month={Oct},    
    year={2024},
    urldate={2025-01-20},
	url={https://www.mckinsey.com/industries/technology-media-and-telecommunications/our-insights/ai-power-expanding-data-center-capacity-to-meet-growing-demand}
}

@article{belkhir2018assessing,
  title={Assessing ICT global emissions footprint: Trends to 2040 \& recommendations},
  author={Belkhir, Lotfi and Elmeligi, Ahmed},
  journal={Journal of cleaner production},
  volume={177},
  pages={448--463},
  year={2018},
  publisher={Elsevier}
}

@article{knowles2022our,
  title={Our house is on fire: The climate emergency and computing's responsibility.},
  author={Knowles, Bran and Widdicks, Kelly and Blair, Gordon and Berners-Lee, Mike and Friday, Adrian},
  journal={Communications of the ACM},
  volume={65},
  number={6},
  pages={38--40},
  year={2022},
  publisher={ACM New York, NY, USA}
}

@techreport{shehabi2024,
  title={{2024 United States Data Center Energy Usage Report}},
  author={Shehabi, Arman and Hubbard,  Alex and Newkirk, Alex and Lei, Nuoa and Siddik, Md Abu Bakkar and Holecek, Billie and Koomey, Jonathan and Masanet, Eric and Sartor, Dale and others},
  institution={Lawrence Berkeley National Laboratory},
   number={LBNL-2001637},
  doi={doi.org/10.71468/P1WC7Q},
  year={2024},
  month={dec}
}

@inproceedings{kermond2010introduction,
  title={An introduction to the algebra of random variables},
  author={Kermond, John},
  booktitle={Proceedings of the 47th Annual Conference of the Mathematical Association of Victoria},
  pages={1--16},
  year={2010}
}

@article{whitehead2015LCA,
  title={The life cycle assessment of a UK data centre},
  author={Whitehead, Beth and Andrews, Deborah and Shah, Amip},
  journal={The International Journal of Life Cycle Assessment},
  volume={20},
  pages={332--349},
  year={2015},
  publisher={Springer}
}

@article{guo2025effect,
  title={The Effect of the Network in Cutting Carbon for Geo-shifted Workloads},
  author={Guo, Yibo and Tomlinson, Amanda and Su, Runlong and Porter, George},
  journal={arXiv preprint arXiv:2504.14022},
  year={2025}
}

@inproceedings{sukprasert2024limitations,
author = {Sukprasert, Thanathorn and Souza, Abel and Bashir, Noman and Irwin, David and Shenoy, Prashant},
title = {On the Limitations of Carbon-Aware Temporal and Spatial Workload Shifting in the Cloud},
year = {2024},
isbn = {9798400704376},
publisher = {Association for Computing Machinery},
address = {New York, NY, USA},
pages = {924–941},
numpages = {18},
location = {Athens, Greece},
series = {EuroSys '24}
}

@article{morris2008analysis,
  title={Analysis of the Dot-Com Bubble of the 1990s},
  author={Morris, John J and Alam, Pervaiz},
  journal={Available at SSRN 1152412},
  year={2008}
}
\vspace{-40pt}
\begin{IEEEbiographynophoto}{Wim Vanderbauwhede} received a PhD in Electrotechnical Engineering from the University of Gent, Belgium in 1996. He is Professor in Low Carbon and Sustainable Computing at the School of Computing Science, University of Glasgow. His research focuses on reducing emissions from computing 
and has resulted in 200 refereed conference and journal papers. Before returning to academic research, he worked as a IC Design Engineer and Senior Technology R\&D Engineer in the Microelectronics industry. 
\end{IEEEbiographynophoto}

\end{document}